\documentclass[showpacs,twocolumn,pre]{revtex4}
\usepackage{graphicx, amsmath, amsthm, amssymb}
\newcommand{\be}{\begin{equation}}
\newcommand{\ee}{\end{equation}}
\newcommand{\bea}{\begin{eqnarray}}
\newcommand{\eea}{\end{eqnarray}}
\newcommand{\bwt}{\begin{widetext}}
\newcommand{\ewt}{\end{widetext}}

\begin{document}

\title{Ponderomotive laser channelling and multi-channelling in homogeneous underdense plasma}
\author{N. Naseri$^{1}$, D. Pesme$^{1,2}$, W. Rozmus$^{1}$ \\
\textit {$^1$Department of Physics, University of Alberta,
   Edmonton, Alberta, Canada } \\
\textit{$^2$Centre de Physique Th{\'e}orique, {\'E}cole Polytechnique, 91128 Palaiseau Cedex,
France}\\
\footnotesize{nnaseri@ualberta.ca }} 
\date{\today}
\maketitle
\section{Introduction}
Propagation of intense laser pulses in  large scale underdense plasma has been an important subject for advanced applications such as inertial confinement fusion \cite{Tabak1994},  particle acceleration \cite{Mangles2005} and radiation sources. The interesting and basic nonlinear physics of relativistic self-focusing (RSF) and self-channelling of intense laser pulses has been also a topic of ongoing theoretical, experimental and simulation studies \cite{Askaryan1962,Max1974,Chen1993,Kurki1989,Feit1998,Tzeng1998,Borisov1998,Pukhov1999,Borghesi1997,Najmudin2003,Esarey2000,Davis2005,Cattani2001,Kim2002,Sun1987,Li2008,Naseri2010}. Relativistic self-focusing happens when the laser power, P, exceeds the critical power $P_{cr}=17 \omega_{0}^{2}/\omega_{p}^{2}$ GW, where $\omega_{0}$ is the laser frequency and $\omega_{p}$ is the plasma frequency\cite{Max1974}. For intense laser pulses the transverse ponderomotive force can be large enough to expel electrons from the central region leading to  full electron evacuation \cite{Sun1987,Feit1998,Cattani2001,Kim2002}.  This problem was first studied by Sun \textit{et. al} \cite{Sun1987} in cylindrical geometry. Later Feit \textit{et. al}\cite{Feit1998} showed that charge conservation was not satisfied in their model. Cattani \textit{et. al} \cite{Cattani2001} and Kim \textit{et. al} \cite{Kim2002} studied this problem in 2D slab geometry and 3D cylindrical geometry, respectively. Their models satisfy the charge conservation by explicitly including Poisson's equation.  In our previous work, we studied channeling of intense laser pulses in underdense plasmas using 2D PIC code in large-scale plasmas.  We have found very good agreement between a stationary model \cite{Cattani2001} and PIC simulations of the laser pulse channeling. We have also observed that for higher plasma densities hosing and transverse instability are more likely to disrupt laser pulse cavitation in ion channels \cite{Naseri2010}.  \newline
Realistic 3D geometry is very important for RSF and self channeling of the laser pulses in underdense plasma and it leads to fundamental differences compared to 2D cases. 3D simulations by Pukhov \textit{et. al}  \cite{Pukhov1999} showed that self focusing is much stronger in 3D than in 2D slab geometry. The goal of this paper is to study channeling, ring structure and stability of the channels using 3D PIC codes Mandor \cite{Romanov2004} and SCPIC \cite{Popov2009}. We will show that azimuthal perturbations can break the symmetry of the nonlinear channel and split the laser pulse into several filaments. The growth length of azimuthal instability is shorter for higher plasma densities. In what follows, first we review the theory of channeling in cylindrical geometry (Sec. \ref{sec4-2},\ref{sec4-3}). Then we illustrate the simulation results for single channelling and compare them with analytical solutions (Sec. \ref{sec4-4}). Section \ref{sec4-6} contains theoretical solutions for the ring structure. The results of simulations and the comparison with analytical solution are presented in Sec. \ref{sec4-7}. Stability of single channels and ring structure is studied in Sec. \ref{sec4-8}. 

\section{A Theoretical Model}\label{sec4-2}
The focus of our work is on the 3D PIC simulations of relativistic laser pulses interacting with a plasma under conditions of complete electron evacuation where wakefields and scattering instabilities are not important. We will study RSF and channeling of short laser pulses using PIC simulations and examine first whether and under which conditions these numerical results approach the stationary solutions of a simplified theoretical model \cite{Gorbunov1973,Sun1987,Chen1993,Kim2002}. This well established theoretical model includes the description of a cold, relativistic electron fluid interacting with a laser pulse that is described in the paraxial approximation. Assuming immobile ions, the basic equations of this model are:
\begin{equation}\label{eqn4-1}
\mathbf{\nabla}^{2}\mathbf{A}-\frac{1}{c^2} \frac{\partial^2\textbf{A}}{\partial t^2}=\frac{1}{c}\frac{\partial}{\partial t}
\mathbf{\nabla}\varphi + \frac{4\pi}{c}Ne\textbf{v},
\end{equation}
\begin{equation}\label{eqn4-2}
\mathbf{\nabla}^{2}\varphi= 4\pi e(N-N_{0}),
\end{equation}
\begin{equation}\label{eqn4-3}
\frac{\partial \textbf{p}}{\partial t}+(\textbf{v}.\nabla)\textbf{p}=\frac{e}{c}\frac{\partial \textbf{A}}{\partial t}+e\nabla \varphi-\frac{e}{c} \textbf{v} \times(\nabla \times \textbf{A}),
\end{equation}
where $N$ is the electron density, $N_{0}$ is the background density, $-e$ is electron charge, $m$ is mass of electron, $\textbf{A}$ is an electromagnetic vector potential and $\varphi$ is an electrostatic scalar potential. Using the equality
\begin{equation}\label{eqn4-4}
(\textbf{v}.\nabla)\textbf{p}=mc^{2}\nabla \gamma-\textbf{v}\times(\nabla \times \textbf{p}),
\end{equation}
we can rewrite (\ref{eqn4-3}) as the equation for the canonical momentum $\textbf{P}_c = \textbf{p}-\frac{e}{c} \textbf{A}$,
\begin{equation}\label{eqn4-41}
\frac{\partial \textbf{P}_c}{\partial t}-\textbf{v} \times (\mathbf{\nabla} \times \textbf{P}_c) = -mc^2 \mathbf{\nabla} \gamma + e\mathbf{\nabla} \varphi ,
\end{equation}
Operating with $\mathbf{\nabla} \times$ on Eq. (\ref{eqn4-41}) we derive the equation for the generalized vorticity, ${\bf \Omega}=\mathbf{\nabla} \times \textbf{P}_c$, 
\begin{equation}\label{eqn4-42}
\frac{\partial \bf{\Omega}}{\partial t}- \mathbf{\nabla} \times (\textbf{v} \times \bf{\Omega})=0.
\end{equation}
It can be shown \cite{Chen1993,Kaganovich2001} using Eq. (\ref{eqn4-42}) that if $\bf{\Omega}=0$ everywhere at some initial time then $\bf{\Omega}$ remains zero at later times. The generalized vorticity vanishes initially in our cold electron plasma in the absence of a laser pulse. Because of $\mathbf{\nabla}\times(\textbf{p}-\frac{e}{c} \textbf{A})=0$ we can introduce a scalar function $\psi$ and write
\begin{equation}\label{eqn4-5}
\gamma m \textbf{v}=\frac{e}{c}\textbf{A}+\mathbf{\nabla}\psi,
\end{equation}
\begin{equation}\label{eqn4-6}
\frac{\partial \psi}{\partial {t}}=e\varphi - mc^{2}(\gamma-1),
\end{equation}
where $\gamma$ is the relativistic factor. The laser pulse is short enough to neglect the ion motion, and we assume that the plasma will approach a quasi-stationary and homogeneous state where $\psi=0$. We introduce the slowly varying, normalized amplitude, $a(r,z)$,  of the vector potential $e\textbf{A}/mc^{2}=(1/2) a(r,z) \exp(i(k_{L}z-\omega t))(\mathbf{e}_{x}+i\mathbf{e}_{y})+c.c$ where $x$ is the direction of propagation and $k_{L}$ is the vacuum wave number. Then the system of Eqs. (\ref{eqn4-1}-\ref{eqn4-6}) will take the following form:
\begin{equation}\label{eqn4-7}
2 i k_{L} \frac{\partial a}{\partial x}+ \mathbf{\nabla_{\perp}}^2  a-\frac{k_{L}^{2}n_{0}}{\gamma}na=0,
\end{equation}
\begin{equation}\label{eqn4-8}
{\nabla_{\perp}}^{2}(\phi)=k_{L}^{2}n_{0}(n-1),
\end{equation}
\begin{equation}\label{eqn4-9}
\phi=\gamma-1, 
\end{equation}
where $\gamma=\sqrt{1+a^{2}}$ is the relativistic factor, $\phi=e \varphi/mc^2$, $n_{0}=N_{0}/n_{c}$, $n=N/N_{0}$ and $n_{cr}$ is the critical density. 
\section{Channelling}{\label{sec4-3}}
RSF at  high laser powers leads to full evacuation of electrons by the ponderomotive force that is enhanced by the relativistic effect and results in the formation of plasma channels. This regime of the nonlinear laser pulse propagation has been of particular interest to the fast ignition scheme, particle acceleration and radiation generation, particularly as one can find the range of laser powers where such channels exhibit a linear stability. In order to find an analytical, or almost an analytical solution describing the cylindrical plasma channel we assume the amplitude of the vector potential in the following form $a(r,z)=a(r)\exp(-i\kappa z)$ where $\kappa$ is the propagation constant. Equations (\ref{eqn4-7}-\ref{eqn4-9})  can be now written as the set ordinary differential equations \cite{Kim2002} in $r$:
\begin{equation}\label{eqn4-10}
{\nabla_{\perp}}^{2}a(r)+(\kappa-\frac{n}{\gamma})a(r)=0,
\end{equation}
\begin{equation}\label{eqn4-11}
{\nabla_{\perp}}^{2}\phi=(n-1),
\end{equation}
\begin{equation}\label{eqn4-12}
\phi=\gamma-1,
\end{equation}
where $\mathbf{\nabla_{\perp}^{2}}=\frac{1}{r}\frac{d}{dr}(r\frac{d}{dr})$ and spatial variables $r$, $z$ are now normalized to $(k_{L}\sqrt{n_{0}})^{-1}$ and $(k_{L}n_{0}/2)^{-1}$ respectively ($r$ and $z$ denote from now on  dimensionless variables). Equations (\ref{eqn4-11}), (\ref{eqn4-12}) give the plasma density $n=1+\nabla_{\perp}^{2} \gamma$. Clearly in this model a strong ponderomotive force can produce full electron expulsion corresponding to $n=0$. For even stronger ponderomotive pressure, which cannot be balanced by the electrostatic force due to charge separation, Eqns. (\ref{eqn4-11}), (\ref{eqn4-12}) can lead to nonphysical negative electron density. To deal with this problem Sun {\it et al}. \cite{Sun1987} suggested a simple solution that amounts to setting $n=0$ inside the cavitation region of the channel whenever $1+\nabla_{\perp}^{2} \gamma < 0$. As pointed out by Feit {\it et al}. \cite{Feit1998} the straightforward application of this fix has resulted in the violation of the overall charge neutrality. Only when the global structure of the solution is determined and the charge conservation is used in evaluation of the channel size \cite{Kim2002} one could proceed with the solution to (\ref{eqn4-10}),  (\ref{eqn4-11}), (\ref{eqn4-12}). Results of Ref. \cite{Kim2002} for a single channel evacuation are reviewed below and later compared with PIC simulations.
Our first goal is to find the threshold power for channeling. Exceeding this power gives an electron cavitated region where $n(r \leq R)=0$. The radius of the electron evacuated channel, $R$, can be calculated by considering the balance between the ponderomotive force and charge separation force. Assuming that $R$ is the boundary position, we can write total charge conservation as:
\begin{equation}\label{eqn4-13}
\int_{0}^{+\infty}(1-n)rdr=\int_{0}^{R}rdr+\int_{R}^{+\infty}(1-n)rdr=0,
\end{equation}
where the plasma density $n=1+\nabla_{\perp}^{2} \gamma$. Substituting $n$  into Eq. (\ref{eqn4-13}) gives:
\begin{equation}\label{eqn4-14}
{[\frac{a(r)a'(r)}{\sqrt{1+a(r)^2}}]}_{r=R}=-\frac{1}{2}R.
\end{equation}
To find a solution to Eq. (\ref{eqn4-10}), we need to know the values of ${a(r)|}_{r=R}$ and ${a'(r)|}_{r=R}$. For inside the channel, Eq. (\ref{eqn4-10}) will be:
\begin{equation}\label{eqn4-15}
\frac{1}{r}\frac{d}{dr}(r\frac{da(r)}{dr})+\kappa a(r)=0
\end{equation}
The solution to this equation is a Bessel function of order zero:
\begin{equation}\label{eqn4-16}
a_{ch}(r)=CJ_{0}(\sqrt{\kappa}r),
\end{equation}
$C$ is the value of ${a(r)|}_{r=0}$. At $r=R$ continuity condition for $a(r)$ and $a'(r)$ gives:
\begin{equation}\label{eqn4-17}
a_{ch}(R)=CJ_{0}(\sqrt{\kappa}R),
\end{equation}
\begin{equation}\label{eqn4-18}
a'_{ch}(R)=-C\sqrt{\kappa}J_{1}(\sqrt{\kappa}R),
\end{equation}
Substituting Eqs. (\ref{eqn4-16}-\ref{eqn4-18}) in Eq. (\ref{eqn4-14}) gives,
\begin{equation}\label{eqn4-19}
C^2=\frac{R^2}{8\kappa J_{1}^{2}(\sqrt{\kappa}R)}(1+\sqrt{1+\frac{16\kappa J_{1}^{2}(\sqrt{\kappa}R)}{R^2J_{0}^{2}(\sqrt{\kappa}R)}}.
\end{equation}
Knowing $C$ for a certain $\kappa$ and fixed $R$, $a(R)$ and $a'(R)$ will be known. \\
To find solution in the plasma region $r>R$, we should combine Eq. (\ref{eqn4-11}) and Eq. (\ref{eqn4-12}) as,
\begin{equation}\label{eqn4-20}
n(r)=1+\frac{1}{r}\frac{d}{dr}(r\frac{d\gamma(r)}{dr}),
\end{equation}

Substituting Eq. \ref{eqn4-20} into Eq. (\ref{eqn4-10}) gives:
\begin{eqnarray}
&& a''(r)+\frac{1}{r}a'(r)-\frac{a(r)a'^2(r)}{(1+a^2(r))}+(\kappa(1+a^2(r)) \nonumber\\
&& -\sqrt{1+a^2(r)})a(r)=0. \label{eqn4-22}
\end{eqnarray}
Equation (\ref{eqn4-22}) can be solved numerically using shooting method with $R$ as the shooting parameter. Boundary conditions are: ${a(r)|}_{r=\infty}=0$ and ${a'(r)|}_{r=\infty}=0$. From the shooting method, we find that the complete evacuation does not happen if $\kappa>0.88$. Therefore the threshold power for channeling $(\kappa_{th}=0.88) $ will be:
\begin{equation}\label{eqn4-23}
P_{th}=\int_{0}^{\infty}a_{\kappa_{th}}^2(r)rdr=1.09P_{cr}
\end{equation}
In next Section, we will show the results of different PIC simulations and compare them with theoretical solutions.
 \begin{figure}
\begin{center}
\includegraphics[totalheight=0.6\textheight]{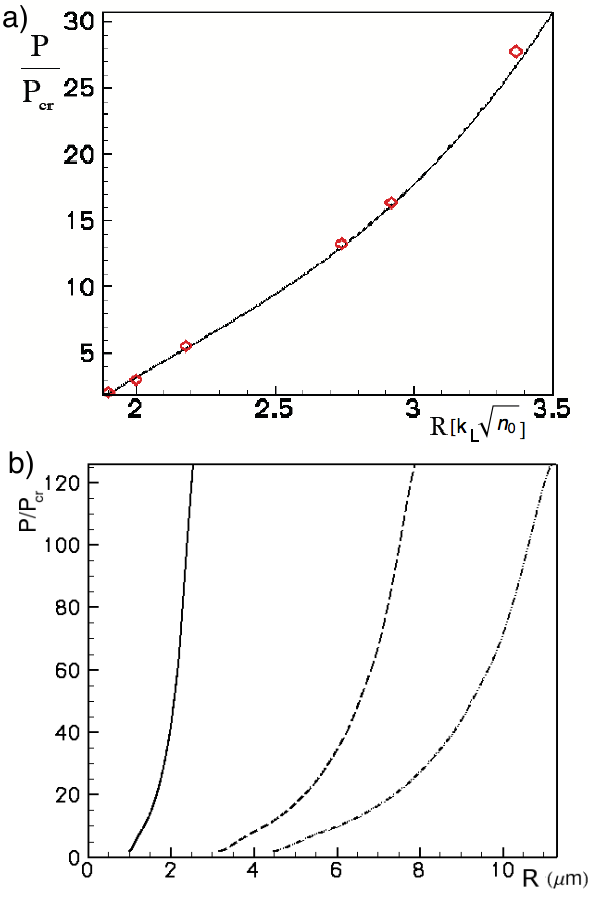}
\end{center}
\caption{$P/P_{cr}$ as a function of evacuated channel radius normalized to $[k_{L}\sqrt{n_{0}}]^{-1}$ (calculated from theoretical model). Points show the simulation results (Fig.1-a).  $P/P_{cr}$ as a function of radius of evacuated channel in $\mu m$ for plasma densities 0.1, 0.01, 0.005 $n_{cr}$ (solid, dashed, dashed-dotted), calculated from theoretical model (Fig.1-b).}\label{fig4-1}
\end{figure}


\section{Single Fully Evacuated Channels-PIC Simulation Results}\label{sec4-4}

\begin{figure}[h]
\begin{center}
\includegraphics[totalheight=0.5\textheight]{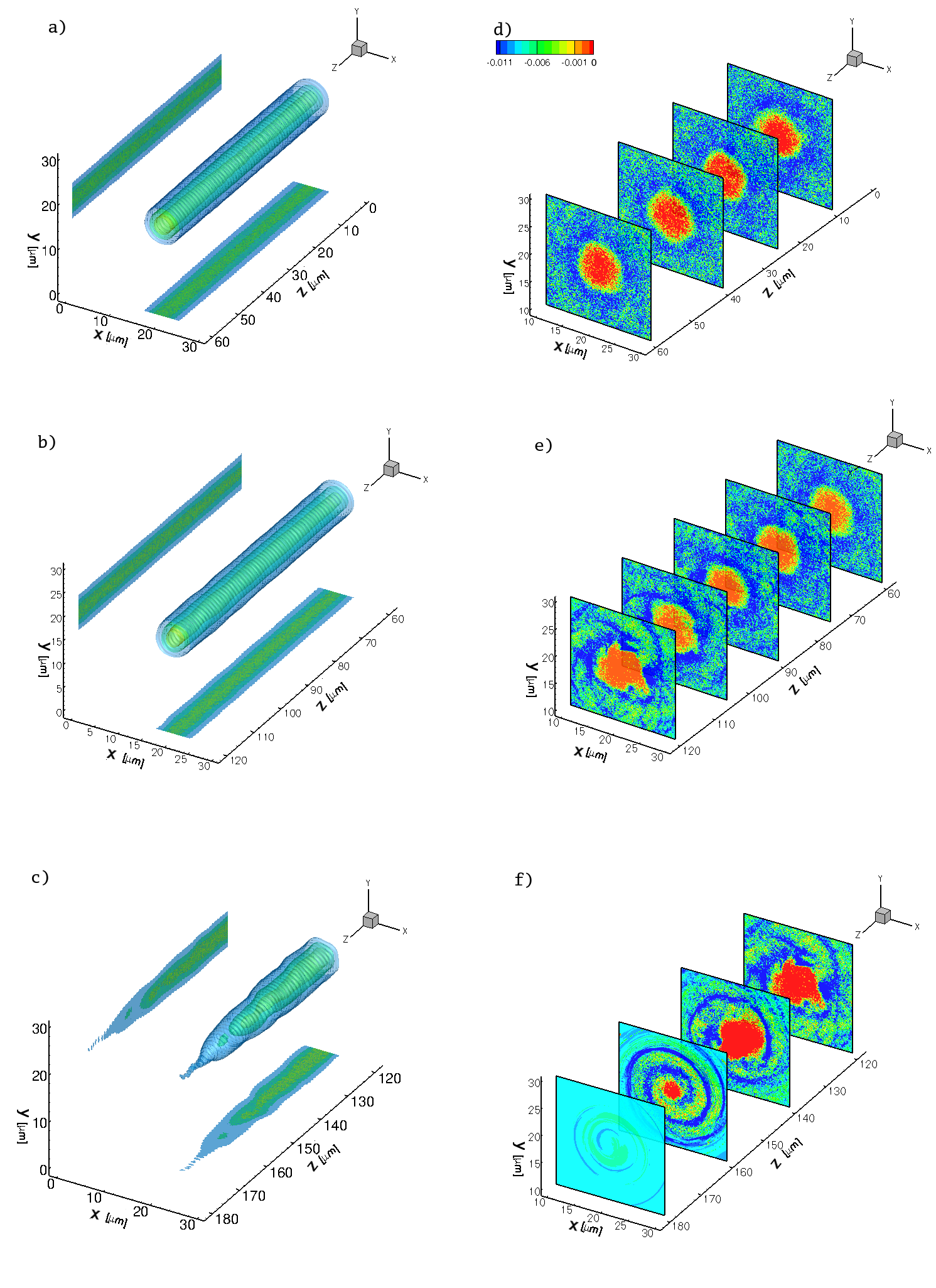}
\end{center}
\caption{(Color online) Contours of laser intensity (a-c) and electron charge density (d-f) normalized to $n_{c}$. }\label{fig4-2}
\end{figure}

\begin{figure}[h]
\begin{center}
\includegraphics[totalheight=0.4\textheight]{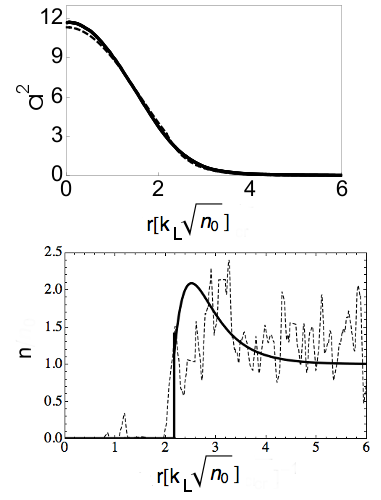}
\end{center}
\caption{Comparison of theoretical solutions  for laser intensity and electron density (solid curves) and 
simulation (dashed curves). Data taken from simulation in z-x plane at z=90 $\mu m$ (t=633 fs).}\label{fig4-3}
\end{figure}

\begin{figure}[h]
\begin{center}
\includegraphics[totalheight=0.4\textheight]{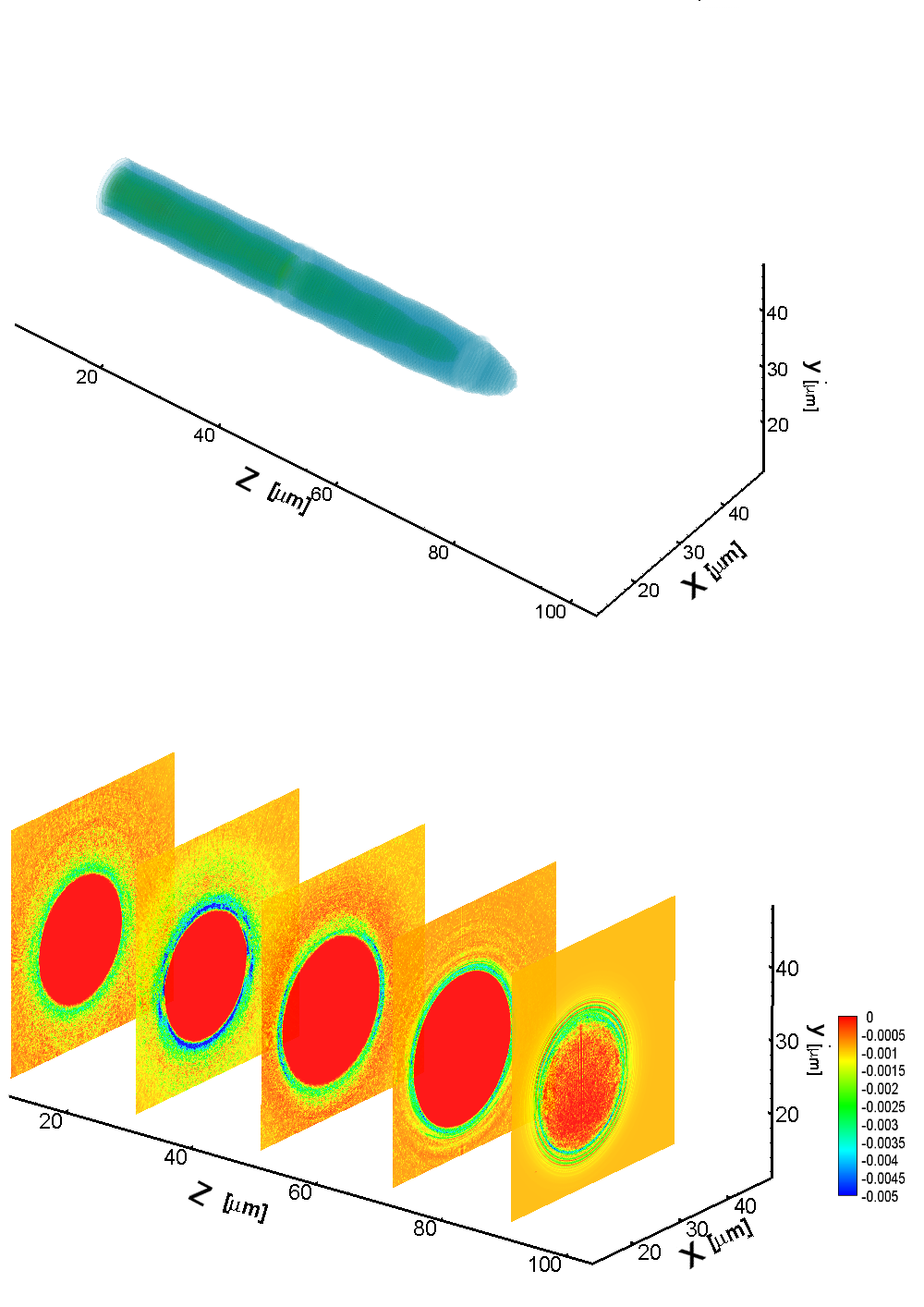}
\end{center}
\caption{(Color online) Contours of laser intensity (a-c) and electron charge density (b-d) normalized to $n_{c}$. }\label{fig4-0-1}
\end{figure}

\begin{figure}
\begin{center}
\includegraphics[totalheight=0.3\textheight]{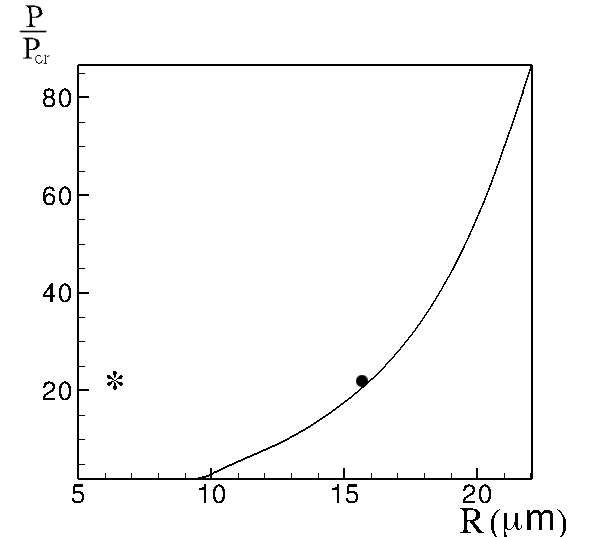}
\end{center}
\caption{$P/P_{cr}$ vs channel radius $R$ in $\mu m$ for plasma density $0.001n_{cr}$ (calculated from theoretical model). Star shows the initial FWHM of the laser intensity. Circle shows the calculated power and channel radius from simulation at $z=90 \mu m$ and t=600 fs. }\label{fig4-0-2}
\end{figure}

We performe several 3D PIC simulations considering different parameters such as different background plasma densities and different laser spot sizes. We have used  PIC codes Mandor \cite{Romanov2004} and SCPIC \cite{Popov2009} in Cartesian geometry. In most of our runs we have used laser pulses with a Gaussian profile, a rise time of $100$ fs and a clamped amplitude afterwards. The pulse is focused $2$ $\mu m$ from the left boundary of the simulation box. The plasma is homogeneous with densities
$(0.001-0.1)n_{cr}$, laser wavelength is $1 \mu m$. The spatial resolution in all simulations that were performed is $\Delta x=\Delta y=\Delta z=\lambda/15$ ($\lambda=1 \mu m$ is laser wavelength). Laser intensity is given by $a^2$, where $a$ is normalized to $a_{0}=m \omega c/e$. The laser pulse is circularly polarized. Light propagates along $z$ direction.  The dimension of simulation box is $z \times x \times y=(200-300) \mu m\times 40 \mu m \times 40 \mu m$. We have calculated the ion plasma period for different plasma densities in simulations and we limited the propagation distance accordingly except for higher plasma densities such as $0.1n_{cr}$ where ions are considered mobile.
We find that for low plasma densities, the incident Gaussian laser pulse with a power above threshold power for channeling reaches a quasi-stationary state which compares well with the solution from Sec. {\ref{sec4-3}}. 
A location of such solutions is illustrated in Fig. \ref{fig4-1}-a (solid curve). Different PIC simulations are shown by dots on the curve. We  find that the radius of evacuated channel is to a very good approximation given by a scaling relation, $R(\mu m)=0.245 \lambda (\mu m)(P/P_{cr})^{(1/4.383)}\sqrt{n_{0}^{-1}}$, where $\lambda$ is the laser wavelength. Having the laser power and the initial plasma density, the radius of the electron free channel can be estimated. It can be seen from this scaling relation that the radius of the channel is smaller for higher plasma densities. Moreover, for the same plasma density, increasing the laser power leads to bigger channel radius. Dimensional graphs of $P/P_{cr}$ as a function of channel radius $R$ $(\mu m)$ helps toward a better understanding of the problem.
Figure \ref{fig4-1}-b shows the total power as a function of channel radius in $\mu m$ for three plasma densities of 0.1, 0.01, 0.005 $n_{cr}$. We see that for the same total power we can get different channel radii,  smaller channels for higher plasma densities.
\newline
Here we discuss the two sample simulations in more details. Figure (\ref{fig4-2}) shows PIC simulation results where the homogeneous plasma density is $0.01n_{cr}$, the initial peak laser intensity is $2.5 \times 10^{19} W/cm^2$ and the initial full-width-half-maximum of the laser intensity is 6  $\mu m$.
Figure \ref{fig4-2} shows the contour plots of laser intensity (a-c) and electron charge density (d-f) after propagating for 200  $\mu m$. After the transient perturbations of the electron density and the laser intensity in front of the laser pulse which corresponds to the rising time of the pulse intensity, the simulation shows a stationary profile. We see in  Fig. \ref{fig4-2}a-b that the laser pulse evacuates electrons, making a cylindrical uniform channel. Figure \ref{fig4-2} c shows the front of the laser pulse.
The density contour plot Fig. \ref{fig4-2} d,e show a straight evacuated channel.  Figure \ref{fig4-2} f shows the interaction of front of the laser pulse with the plasma. The characteristic spiral shape in the electron density is due to the circular polarization of the laser pulse.  
We show the calculated analytical and simulation result for this simulation in Fig. \ref{fig4-3}. The data from the simulation is taken from x-y plane at x=90  $\mu m$. We calculated the power of the laser, which is $P/P_{cr}=5.0$ close to the expected value from theoretical solution $P/P_{cr}=5.5$.\\
For the incident Gaussian pulse which has an initial power and a spot size different from the parameters of  the stationary solutions in Fig. \ref{fig4-1}, the evolution towards these asymptotic states has involved focusing or defocusing of the laser pulse until it reaches the right width of the stable channel. Figures \ref{fig4-0-1} show another simulation results where the homogeneous plasma density is $0.001n_{cr}$, the peak laser intensity is $5.0 \times 10^{20} W/cm^2$ and the initial full-width-half-maximum of the laser intensity is $6$ \hspace{1 mm} $\mu m$. Figures \ref{fig4-0-1}a,c show the contour plots of intensity profile of the laser and Figs. \ref{fig4-0-1}b,d the density profile of electrons. The ponderomotive force pushes the electrons away from regions with higher laser intensity making a completely evacuated channel. The diameter of the evacuated channel is  $\sim 30 \mu m$. The laser power is $\sim 22 \times P_{cr}$. In this example, the initial laser spot size is smaller than the expected channel radius (Figs. \ref{fig4-0-2},), therefore the laser pulse first adjusts itself,  broadens, increasing the spot size and then the channel becomes stable, (see Figs. \ref{fig4-0-2}). The solid curve in Fig. \ref{fig4-0-2} shows the $P/P_{ch}$ vs full channel width $R$ for plasma density $0.001n_{cr}$. The star represents the input power and the initial full-width-half-maximum of the laser intensity. The  circle represents the calculated power and channel width taken from transverse lineout in the simulation box. 

\section{Ring Modes}\label{sec4-6}
\subsection{Theoretical Review}\label{subsec5-1}
\begin{figure}[h]
\begin{center}
\includegraphics[totalheight=0.25\textheight]{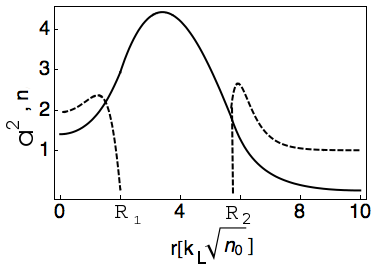}
\end{center}
\caption{Schematic graph showing the ring structure solution. The solid curve is the amplitude of the laser and the dashed curve is the electron density calculated for $\kappa=0.3$.  }\label{fig4-6}
\end{figure}

Central depleted channels are not the only solutions to Eqs. (\ref{eqn4-10}-\ref{eqn4-12}). A central electron filament enclosed by  an evacuated ring is another possible set of solutions to Eqs. (\ref{eqn4-10}-\ref{eqn4-12})\cite{Kim2002}. These structures can exist at higher laser powers. To solve the equations,  the shooting parameter will be the on-axis value of the field and its first derivative on axis will be zero because of symmetry. There is a freedom in choosing the boundary position ($R_{1}$, Fig. \ref{fig4-6}). The only constraint is that the electron density can not be negative. The minimum power for this structure for specific $\kappa$ is when the electron density at $R_{1}$ is zero. Having $R_{1}$, we know the field amplitude and its first derivative at this point. \\
The solution for the depleted region (from $R_{1}$ to $R_{2}$) is:
\begin{equation}
a(r)=A_{v} J_{0}(\sqrt{\kappa }r)+B_{v} Y_{0}(\sqrt{\kappa }r),
\end{equation}
where $J_{0}$ and $Y_{0}$ are zero order Bessel and Neumann functions respectively. Therefore the solution at the boundary is:
\begin{equation}
a(R_{1})=A_{v}J_{0}(\sqrt{\kappa} R_{1})+B_{v} Y_{0}(\sqrt{\kappa} R_{1}),
\end{equation}
\begin{equation}
a'(R_{1})=-A_{v} \sqrt{\kappa} J_{1}(\sqrt{\kappa} R_{1})-B_{v} \sqrt{\kappa}Y_{1}(\sqrt{\kappa} R_{1}).
\end{equation}
Integrating Poisson's equation:
\begin{eqnarray}
&& \int_{0}^{+ \infty}(1-n)r dr=
 \int_{0}^{R_{1}}(1-n) r dr+  \nonumber \\
&& \int_{R_{1}}^{R_{2}} r dr
+\int_{R_{2}}^{+ \infty}(1-n) r dr=0,
\end{eqnarray}
gives a relation between $R_{1}$ and $R_{2}$:
\begin{equation}
g(R_{1})-g(R_{2})=\frac{1}{2}(R_{2}^{2}-R_{1}^{2}),
\end{equation}
where
\begin{equation}
g(r)=\frac{r a(r) a'(r)}{\sqrt{(1+a^{2} (r)}}.
\end{equation}
Therefore $R_{2}$ can be calculated from above equations. Knowing $R_{2}$, we can calculate $a(R_{2})$ and $a'(R_{2})$ and the solution for the semi-infinite plasma region can be found.     
The threshold power for exciting ring structure solution for fixed $\kappa$ is when the density at the first boundary from central axis becomes zero. Figure \ref{fig4-7-1} shows the minimum total power required  for ring structure as a function of $\kappa$. The minimum power to excite these structures is $33P_{th}$.  

\subsection{PIC Simulations}\label{sec4-7}
\begin{figure}[h]
\begin{center}
\includegraphics[width=0.55\textwidth]{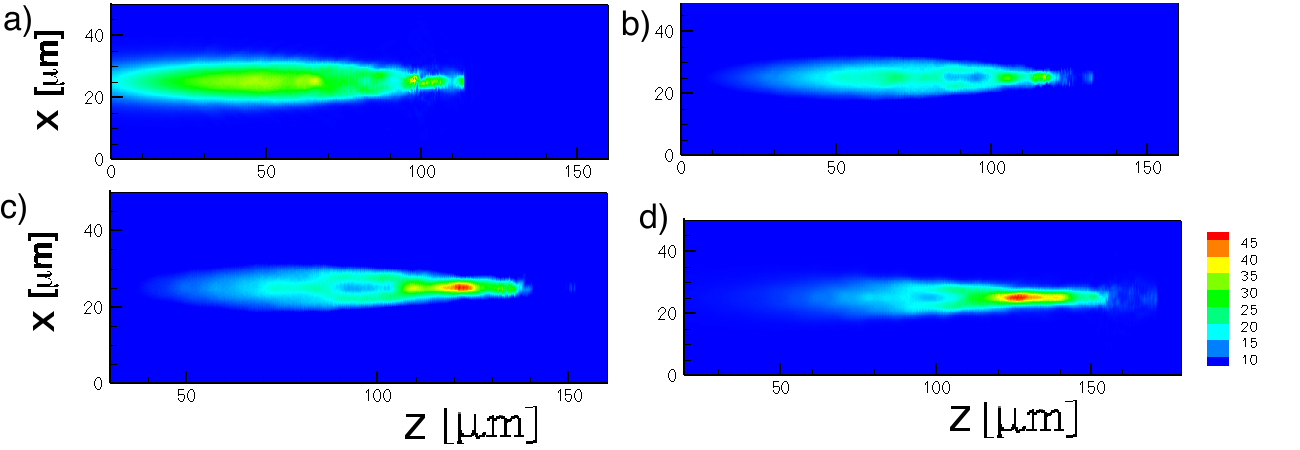}
\end{center}
\caption{(Color online) Contour plots of laser intensity at different times. }\label{fig4-6-1}
\end{figure}

\begin{figure}[h]
\begin{center}
\includegraphics[totalheight=0.5\textheight]{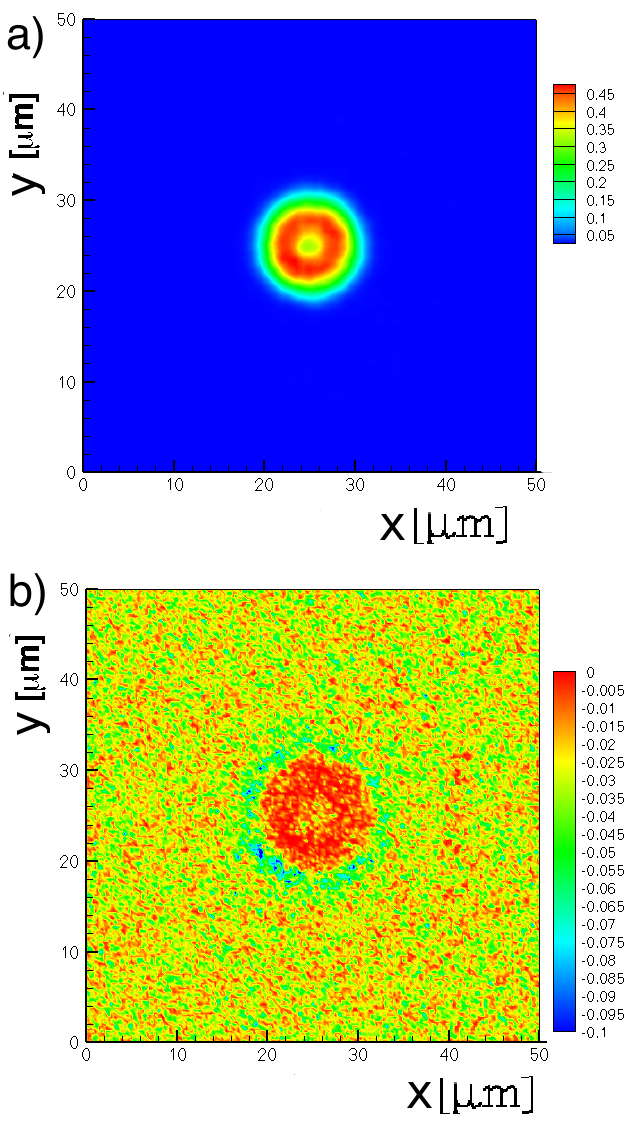}
\end{center}
\caption{(Color online) Contour plots of laser intensity (a) and electron charge density (b) in x-y plane at z=95 $\mu m$ and t=400 fs. }\label{fig4-6-2}
\end{figure}

\begin{figure}[h]
\begin{center}
\includegraphics[totalheight=0.4\textheight]{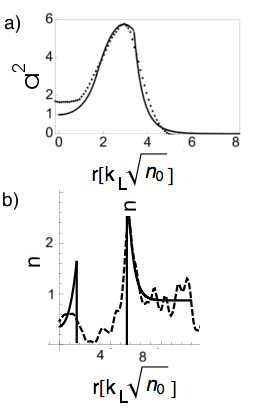}
\end{center}
\caption{Amplitude of the laser  (a) and electron density (b), solid curves are the theoretical solutions and dashed curves are taken from simulation in x-x plane at z=95 $\mu m$.  }\label{fig4-6-3}
\end{figure}

Ring structures are observed in our 3D simulations.   As we input the laser power into the simulations in the form of the Gaussian pulse, we always observe ring modes coexisting with the main laser mode. We will  present the results of a sample simulation with 3D PIC code SCPIC \cite{Popov2009}. The input parameters are as follows: the initial homogeneous plasma density is 0.036$n_{cr}$. Initial peak laser intensity is $10^{20}$ $W/cm^{2}$ and FWHM of the Gaussian laser pulse is $10$ $\mu m$ ($P \sim 230 P_{th}$). The FWHM of the laser pulse  is 250 fs and is focused 10 $\mu m$ from the left boundary. Figure  \ref{fig4-6-1} shows the  contour plots of time evolution of the laser intensity in the xy plane. In the beginning we see the formation of a single channel. Then the laser pulse  starts splitting into two from the front after propagating for 100 $\mu m$. The formation of uniform ring  is illustrated  in Fig. \ref{fig4-6-2} which shows the laser intensity and electron charge density in y-z plane at $z=95 \mu m$. Figure \ref{fig4-6-3} shows the comparison between theoretical and simulation results for ring structure ($\kappa=0.3$ ). 
At later time, the ring structure goes back in the pulse. This is because the coexisting main mode has smaller $\kappa$, therefore has a larger group velocity compared to the ring structure. Thus the ring structure is slower than the main mode and moves back in the pulse. 

\section{Stability Analysis }\label{sec4-8}
\subsection{Stability Analysis-Single Channel}\label{subsec7-1}
\begin{figure}[h]
\begin{center}
\includegraphics[width=0.45\textwidth]{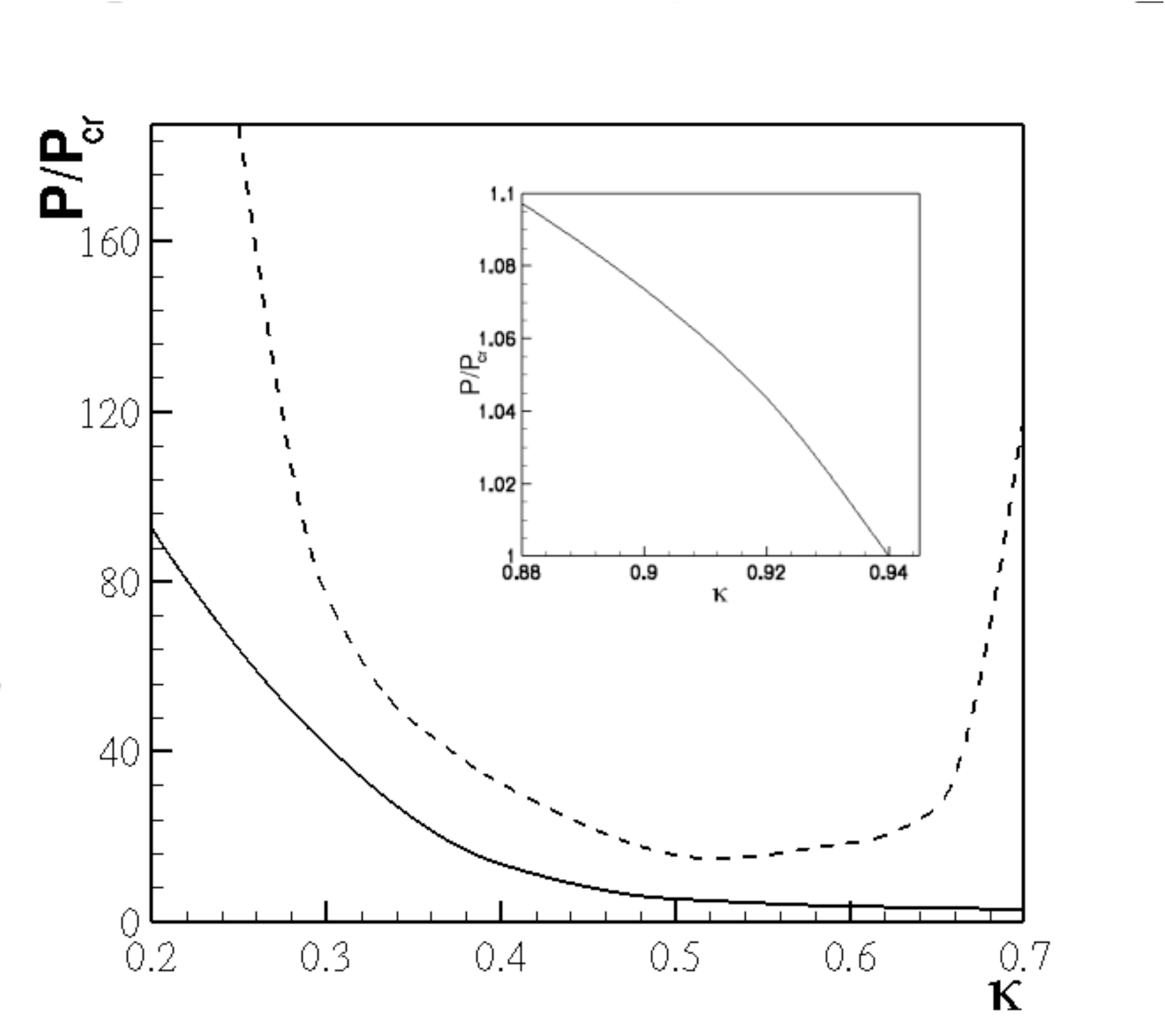}
\end{center}
\caption{The solid curve shows the power as a function of $\kappa$ for single channel solution. The dashed curve shows the minimum power required to obtain ring structure as a function of $\kappa$. The total power as a function of $\kappa$ for partial evacuation is shown in the inset.}\label{fig4-7-1}
\end{figure}

We will examine the stability of theoretical solutions to Eqs. (\ref{eqn4-10}-\ref{eqn4-12}) by linear stability analysis. Within the stationary approximation of Sec\ref{sec4-2} we rewrite Eq. (\ref{eqn4-10})  as:
\begin{equation}\label{eqn4-7-1}
i \frac{\partial a}{\partial z}+\nabla_{\perp}^2 a-\frac{na}{\gamma}=0,
\end{equation}
where $r$ is normalized to $[k_{L}\sqrt{n_{0}}]^{-1}$ and $z$ is normalized to $[k_{L}n_{0}/2]^{-1}$. To investigate the stability of solutions to Eqs. (\ref{eqn4-10}-\ref{eqn4-12}), we assume an exponentially growing small perturbation as $a=a_{0}(r)+a_{1}(r)\exp(\delta z)$ where $a_{0}$ is the theoretical solution to Eqs. (\ref{eqn4-10}-\ref{eqn4-12}) and $a_{1}=u+iv$,  substituted into Eq. (\ref{eqn4-7-1}) upon linearization leads to: 
\begin{equation}\label{eqn4-7-2}
\delta u=-L_{0} v, \hspace{2mm} -\delta v=L_{1} u.
\end{equation}
The growth rate of small perturbation is determined by the above eigenvalue problem where the operators $L_{0}$ and $L_{1}$ are defined as:
\begin{equation}\label{eqn4-7-3}
L_{0}=\kappa-\frac{1}{\gamma_{0}}+\frac{1}{r\gamma_{0}^2}\frac{d}{dr}+\frac{1}{\gamma_{0}^2}\frac{d^2}{dr^2}-\frac{a_{0}'^2}{\gamma_{0}^4},
\end{equation}
\begin{equation}\label{eqn4-7-4}
L_{1}=L_{0}+\frac{a_{0}^2}{\gamma_{0}^3}-\frac{2a_{0}a_{0}'}{r\gamma_{0}^4}-\frac{a_{0}a_{0}''}{\gamma_{0}^4}-\frac{2a_{0}a_{0}'}{\gamma_{0}^4}+\frac{4a_{0}^2a_{0}'^{2}}{\gamma_{0}^6},
\end{equation}
where  $\gamma_{0}=\sqrt{1+a_{0}^2}$ and $L_{0}a_{0}=0$. Therefore 
\begin{equation}
L_{0}L_{1}u=-\delta^2u.
\end{equation} 
It can be shown that  a sufficient condition for the stability of solutions to Eqs. (\ref{eqn4-10}-\ref{eqn4-12}) against symmetric perturbation is  $\frac{\partial P}{\partial \kappa}<0$ \cite{Vakhitov1973}. With reference to Fig. \ref{fig4-7-1} we see that both partially evacuated channel solution ($\kappa > 0.88$) and fully evacuated channel solution ($\kappa \le 0.88$) are stable against symmetric perturbations\cite{Kim2002}. 
\subsection{Stability Analysis (symmetric perturbation)-Ring Structure}\label{subsec7-2}
Figure \ref{fig4-7-1} shows the total power as a function of propagation constant $\kappa$ for ring structure solution when $n_{R_1}=0$ (dashed curve). As mentioned earlier the solutions with $n_{R1}=0$ gives the least power required for ring structure formation\cite{Kim2002}.  For  $\kappa \leq 0.5$,  the stability condition $\frac{\partial P}{\partial \kappa}<0$ holds. This means that the ring structure solution for $\kappa \leq 0.5$ is stable against  symmetric perturbations. However for  $\kappa > 0.5$, the stability condition is not satisfied $\frac{\partial P}{\partial \kappa}>0$\cite{Kim2002} . Therefore ring structure solution with propagation constant $\kappa > 0.5$ is unstable against symmetric perturbation.  We should mention that  the above discussion is for a special case of $n_{R1}=0$. For other possible cases where $n_{R1} \neq 0$, one has to plot the corresponding graph and examine the stability condition $\frac{\partial P}{\partial \kappa} <0$.
\subsection{Stability Analysis (asymmetric perturbation)-Ring Structure} \label{subsec7-3}
\begin{figure}[h]
\begin{center}
\includegraphics[totalheight=0.3\textheight]{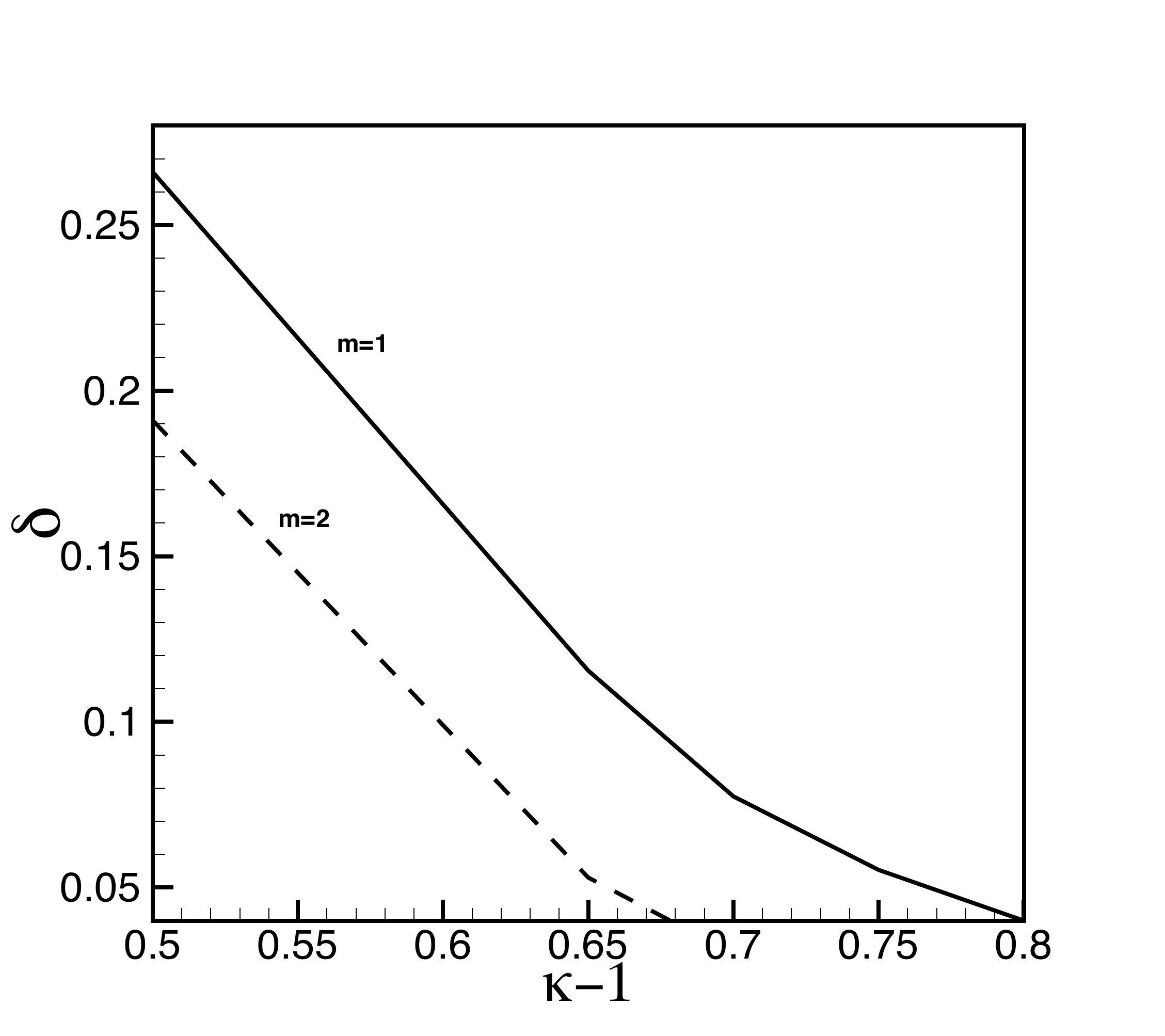}
\end{center}
\caption{Growth rate of transverse instability $\delta$ as a function of $\kappa-1$ for ring structure $m=1,2$. }\label{fig4-7-3}
\end{figure}

In this Section we examine the stability of ring structure for asymmetric (azimuthal index=1,2,...) eigenfunctions.  We will show that the break up of the ring structure can be due to a transverse instability. To investigate the stability of ring structure, we use approximate stability theory \cite{Soto1991,Atai1994} where all quantities are the mean values averaged over annular ring (evacuated region). We assume azimuthally dependent  perturbation as:
\begin{equation}\label{eqn4-7-5}
a=a_{0}+\mu \cos(m\theta) \exp(\delta z),
\end{equation}
where $\mu$ is a small constant and $\delta$ is the growth rate. The form chosen here ensures that the field is periodic in $\theta$ and it allows the modulation to develop around the ring. This means that we have azimuthal perturbation which as will be shown can break the azimuthal symmetry of the ring. Substituting Eq. (\ref{eqn4-7-5}) into Eq. (\ref{eqn4-7-1}) upon linearization leads to:
\begin{equation}
\delta^2=(\kappa-\frac{m^2}{\overline{r}^2})(\kappa-\frac{m^2}{\overline{r}^2\overline{\gamma_{0}}^2}-\frac{\overline{a_{0}} \overline{a''_{0}}}{\overline{\gamma_{0}}^4}+\frac{3 \overline{a_{0}}\overline{a'_{0}}^2}{\overline{\gamma_{0}}^6}-\frac{\overline{a_{0}}\overline{a'_{0}}}{\overline{r}\overline{\gamma_{0}}^4}),
\end{equation}
where $\overline{Q}=\int_{R_1}^{R_2} Q a_{0}^2 r dr/ \int_{R_1}^{R_2}  a_{0}^2 r dr$ with $Q$ which denotes any of the quantities  $a_{0}, a'_{0}, r, \gamma_{0}$. Figure \ref{fig4-7-3} shows the growth rate of transverse instability as a function of  $\kappa-1$. Solid curve shows the growth rate of the transverse instability for $m=1$ and dashed-curve for $m=2$. It can be seen that $m=1$ dominates over $m=2$. The characteristic gain length ($1/ \delta$) can be very long for lower plasma densities.  

\section{Instability of ring modes}\label{sec4-9}
\begin{figure}
\begin{center}
\includegraphics[width=0.38\textwidth]{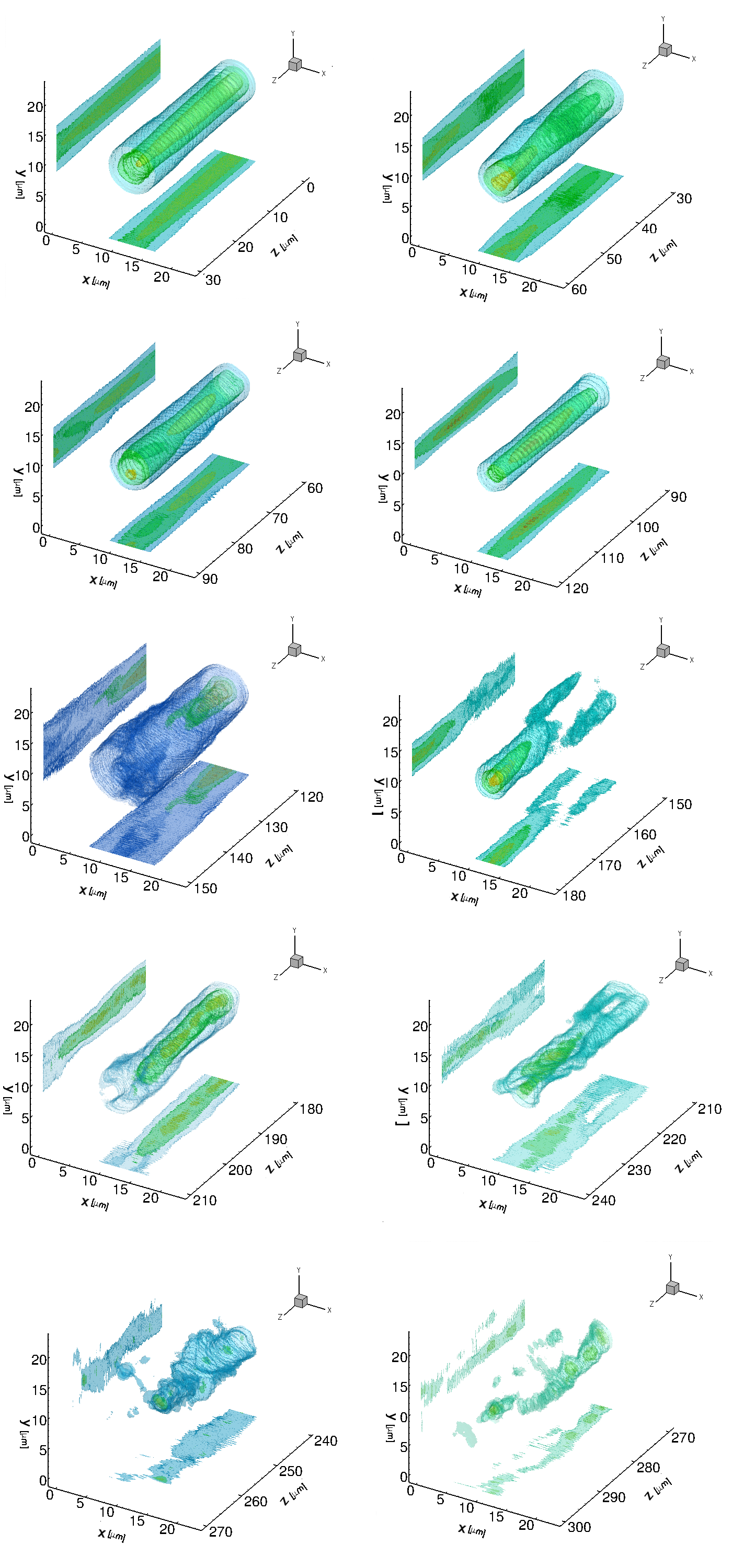}
\end{center}
\caption{(Color online) Contour plots of laser intensity after propagating for 300 $\mu m$. }\label{fig4-8-2}
\end{figure}
Azimuthal perturbations due to the interaction of an intense laser pulse with a plasma can break up the symmetry of the laser pulse as addressed in previous Section. We observed the symmetry breaking of the laser pulse in simulations in higher plasma 
densities ($n_{0} \geq 0.1$).  We could not reach the stable channeling in simulations as predicted from analytical solutions for $n/n_{cr} \geq 0.1$. Azimuthal instability becomes more important in this regime because the gain length is smaller.  

Figure \ref{fig4-8-2} shows a simulation with following parameters: initial plasma density is 0.1$n_{cr}$, the peak laser intensity is $4.7 \times 10^{19}$ $W/cm^{2}$ and the initial FWHM of the laser intensity is $5$ $\mu m$. This simulation was performed twice with both movable and stationary ions. Figure \ref{fig4-8-2} shows the iso-surfaces of evolution of the laser pulse as it propagates through the plasma for 300 $\mu m$. We see the formation of a ring structure after propagating for z=17 $\mu m$. Figure \ref{fig4-8-3} shows a cut at z=17  $\mu m$ in y-z plane. The top panel, bright ring, shows the laser intensity and the lower panel shows the electron charge density. The plasma density follows the same pattern, central electron filament enclosed by a cavitated ring. This structure correspond to $\kappa=0.4$. As the laser pulse propagates farther, the ring loses its symmetrical shape.  

\begin{figure}[h]
\begin{center}
\includegraphics[totalheight=0.25\textheight]{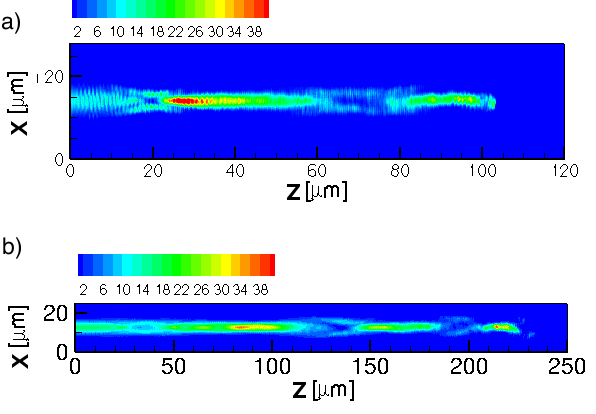}
\end{center}
\caption{(Color online) Contour plots of laser intensity in z-x plane at t=464 and 943 fs. }\label{fig4-8-5}
\end{figure}

Azimuthal perturbation breaks the symmetry of ring structure as it propagates through plasma (See \ref{subsec7-3} ). Figure \ref{fig4-8-5} shows the contour plots of the laser intensity in z-x plane at t=464 and 943 fs. 
Figure \ref{fig4-8-4} shows different cuts at different positions in the y-z plane at z=21, 44 and 70 $\mu m$ at t=580 fs. Figure \ref{fig4-8-4}-a shows a ring structure at z=21 $\mu m$, central maximum at z=44 $\mu m$ and nonuniform ring at z=70 $\mu m$.
The growth rate of the transverse instability for ring structure for this simulation ($\kappa=0.4$) is 0.1 which corresponds to a gain length of 33 $\mu m$ for these parameters.  
\begin{figure}
\begin{center}
\includegraphics[totalheight=0.4\textheight]{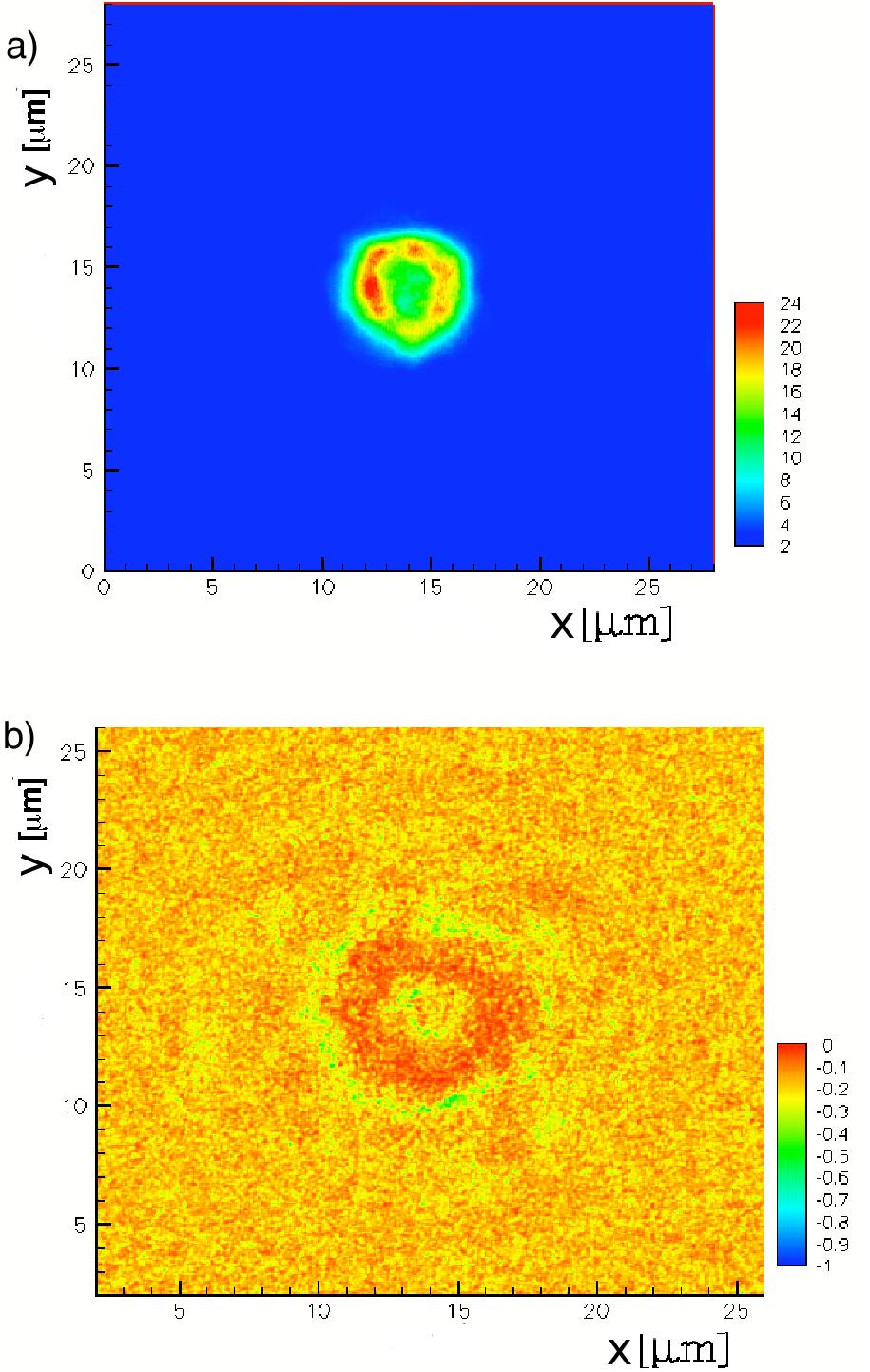}
\end{center}
\caption{(Color online) Contour plots of laser intensity (a) and electron charge density (b) in x-y plane at z=17 $\mu m$ and t=133 fs. }\label{fig4-8-3}
\end{figure}

\begin{figure}
\begin{center}
\includegraphics[totalheight=0.5\textheight]{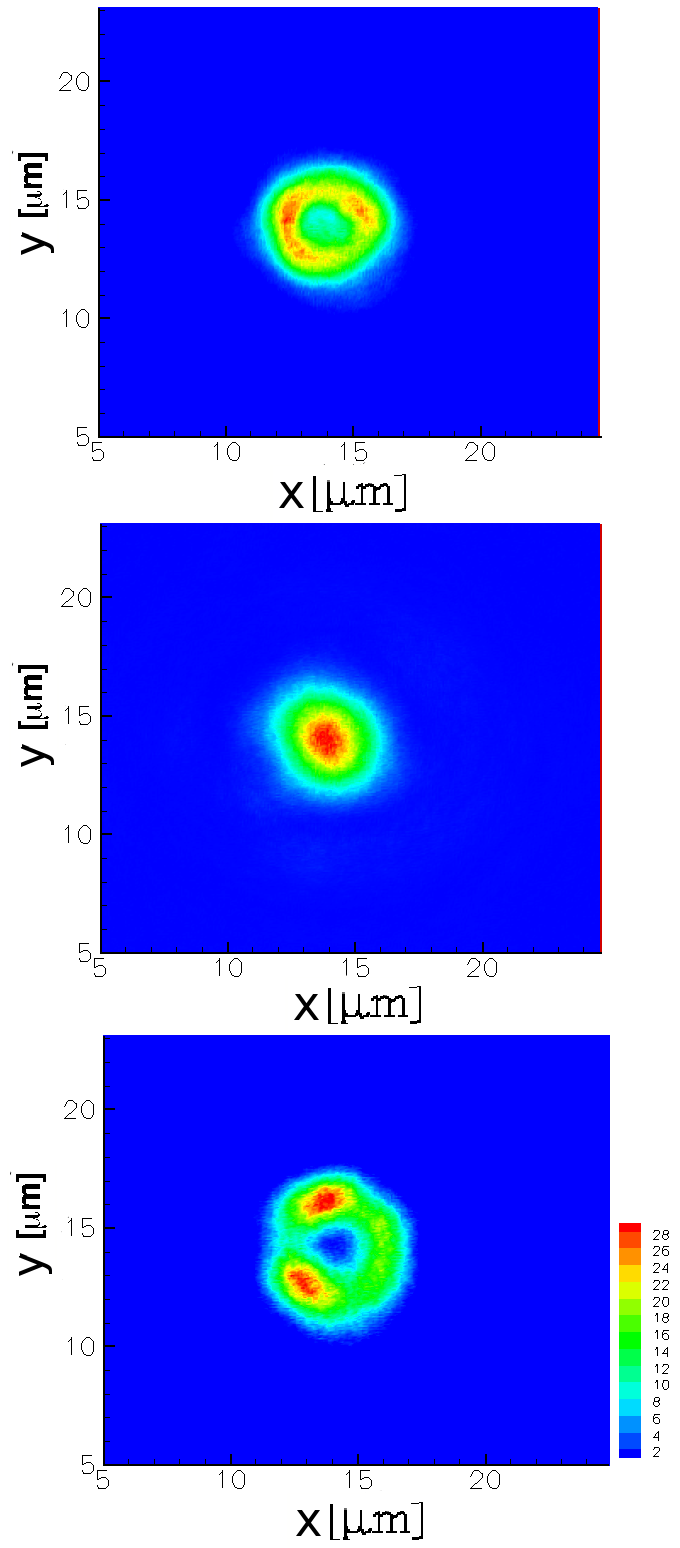}
\end{center}
\caption{(Color online) contour plots of laser intensity showing in x-y plane, cuts at z=21, 44 and 70 $\mu m$ at t=580 fs.
 }\label{fig4-8-4}
\end{figure}  

\section{Conclusions}
We have studied laser pulse channeling in underdense plasma by means of analytical theory and 3D PIC simulations. Numerous theoretical studies over the years \cite{Sun1987,Feit1998,Borisov1998,Kurki1989,Chen1993} and in particular the Ref. \cite{Kim2002} have provided the set of analytical results which are used together with 3D PIC simulations in constructing different scenarios of  laser pulse channeling. 

The most significant result of this paper is the demonstration of the single fully evacuated stationary channel solution can be reached an asymptotic state in PIC simulations. In the density range $0.001<n_{0}<0.1$ and for a laser power above channeling power, $P_{th} \sim 1.1 P_{cr}$  , we were able to reproduce in PIC simulations the analytical curve from Fig. \ref{fig4-1}. It describes the location of stationary analytical solutions in terms of captured power vs channel radius.  We showed that single channels are stable structures against symmetric perturbations. We have not observed stable channeling for plasma densities $\ge 0.1n_{cr}$. 

The excitation of the surface waves on the edges of the fully evacuated channel was addressed. The amplitude of the excited surface waves grows as the laser propagates through the plasma and so does the energy of the electrons on the edges of the evacuated channel.  Eventually the electrons will be heated and start filling the channel. It is important to note that the excitation of the surface waves can happen if the ascending part of the laser is comparable with the surface wave wavelength. 

We also studied the formation of the ring structure in theory and simulations. An evacuated ring enclosed by an electron filament was observed in our 3D simulations. However they always coexist with the main laser mode. The threshold power for ring structure formation is when $P>33P_{cr}$.  Higher laser power is needed for ring formation. These rings are stable against symmetric perturbations if $\kappa<0.5$ (Fig. \ref{fig4-7-3}). Our studies on stability of the rings against asymmetric  perturbations show that ring structure is not stable against azimuthal perturbations. The growth rate of the instability is shorter for higher densities and it is therefore more effective at higher densities.  In fact, we presented an example (Fig. \ref{fig4-8-5}) for the density $0.1n_{cr}$ where the evacuated ring forms early in the simulation, however due to transverse instability the ring collapses and nonuniform rings form(Fig. \ref{fig4-8-4}). 

\end{document}